# PAIRING PROPERTIES OF SUPERHEAVY NUCLEI

A. STASZCZAK[a], J. DOBACZEWSKI[b,c,d], W. NAZAREWICZ[b,c,d]

[a] *Department of Theoretical Physics, Institute of Physics, Maria Curie-Skłodowska University, pl. M. Curie-Skłodowskiej 1, 20-031 Lublin, Poland*

[b] *Department of Physics, University of Tennessee, Knoxville, TN 37996, Knoxville, USA*

[c] *Physics Division, Oak Ridge National Laboratory, P.O.Box 2008, Oak Ridge, TN 37831, USA*

[d] *Institute of Theoretical Physics, Warsaw University, ul. Hoża 69, Warsaw, Poland*

Pairing properties of even-even superheavy $N$=184 isotones are studied within the Skyrme-Hartree-Fock+BCS approach. In the particle-hole channel we take the Skyrme energy density functional SLy4, while in the particle-particle channel we employ the seniority pairing force and zero-range $\delta$-interactions with different forms of density dependence. We conclude that the calculated static fission trajectories weakly depend on the specific form of the $\delta$-pairing interaction. We also investigate the impact of triaxiality on the inner fission barrier and find a rather strong $Z$ dependence of the effect.

## 1. Introduction

In the almost fifty years since the phenomenon of superconductivity was brought into nuclear structure,[1,2,3] our knowledge of the nature of pairing correlations in nuclei is still unsatisfactory and many fundamental questions remain (see Refs.[4,5,6] and papers quoted therein). For instance, some components of the pairing interaction are believed to be induced, like in solid-state superconductors,[7] some are directly rooted in the nucleon-nucleon force. In practical calculations based on effective interactions, one considers various pairing parametrizations. The resulting pairing fields have a strong influence on most low-energy properties of the nuclei[8] and the nuclear large-amplitude collective motion. Indeed, by increasing configuration mixing and reducing the magnitude of symmetry-breaking effects, pairing tends to make the nuclear collective motion more adiabatic,[9] Therefore, when aiming at a quantitative understanding of fission properties of heavy and superheavy elements, it is important to have the pairing channel under control.

The purpose of this study is to compare different pairing schemes that are currently used in the Skyrme-Hartree-Fock+BCS (SHF+BCS) model to describe nuclear superfluidity. Within this framework, we compare the seniority paring force and state-dependent $\delta$-interaction (referred to as SHF+BCS(G) and SHF+BCS($\delta$), respectively). As in our previous paper,[10] we focus here on the fission properties of even-even isotones with $N$=184; namely, we discuss the total binding energies, mass hexadecapole moments, and pairing gaps calculated along the static fission paths.

## 2. Theoretical framework and results

### 2.1. *Description of pairing correlations*

In our SHF+BCS approach, we use the Skyrme energy density functional in its SLy4 parameterization[11] in the particle-hole channel, whereas two different pairing schemes were implemented in the particle-particle channel. The SHF+BCS(G) scheme employs the seniority pairing force with strength parameters defined as in Ref.[12], i.e.,

$$\begin{aligned} G^n &= \left[19.3 - 0.084\,(N-Z)\right]/A\,, \\ G^p &= \left[13.3 + 0.217\,(N-Z)\right]/A\,, \end{aligned} \tag{1}$$

additionally scaled by

$$\tilde{G}^{n/p} = f_{n/p} G^{n/p}. \tag{2}$$

In the SHF+BCS($\delta$) scheme, we apply the state-dependent $\delta$-interaction[13] with commonly used parameterization variants,[14] which are summarized as

$$V_\delta^{n/p}\left(\vec{r}_1,\vec{r}_2\right) = V_0^{n/p}\left[1-\eta\left(\frac{\rho\left(\frac{\vec{r}_1+\vec{r}_2}{2}\right)}{\rho_0}\right)\right]\delta\left(\vec{r}_1-\vec{r}_2\right), \tag{3}$$

where $\rho_0 = 0.16\,\mathrm{fm}^{-3}$ and

$$\eta = \begin{cases} 0\,, & \text{for the delta interaction (DI), } \textit{volume} \text{ pairing;} \\ 1\,, & \text{for the density-dependent delta interaction (DDDI), } \textit{surface} \text{ pairing;} \\ \frac{1}{2}\,, & \text{for the mixed } \textit{volume} \text{ and } \textit{surface} \text{ pairing (MIX).} \end{cases} \tag{4}$$

The scaling factors of Eq. (2), $f_n = 1.41$ and $f_p = 1.13$, and pairing strengths $V_0^n = 282.0$ MeV, $V_0^p = 285.0$ MeV (DI), $V_0^n = 842.0$ MeV, $V_0^p = 1020.0$ MeV (DDDI), and $V_0^n = 425.5$ MeV, $V_0^p = 448.5$ MeV (MIX) were adjusted to reproduce the experimental[15] neutron ($\Delta_n = 0.696$ MeV) and proton ($\Delta_p = 0.803$ MeV) pairing gaps in $^{252}$Fm. As we deal with contact interactions, we use a finite pairing-active space defined by including $\Omega^{n/p} = (N \text{ or } Z)$ lowest single-particle states for neutrons and protons, respectively. In the SHF+BCS($\delta$) approach, the pairing gap is state dependent. Therefore, the average (spectral) gaps,

$$\langle\Delta_{n/p}\rangle = \frac{\sum_{k\in\Omega^{n/p}} v_k u_k \Delta_k}{\sum_{k\in\Omega^{n/p}} v_k u_k}\,, \tag{5}$$

were used as measures of experimental pairing gaps deduced from the odd-even mass staggering. In Eq. (5) $v_k$ and $u_k$ are the BCS occupation amplitudes (see, e.g., Ref.[16] for a more detailed discussion).

The calculations were carried out using the code HFODD (v.2.19l)[17,18,19] that solves self-consistent HF equations by using a Cartesian 3D deformed harmonic-oscillator finite basis. In the calculations, we took the lowest 1140 single-particle states for the basis. This corresponds to 17 oscillator shells at the spherical limit.

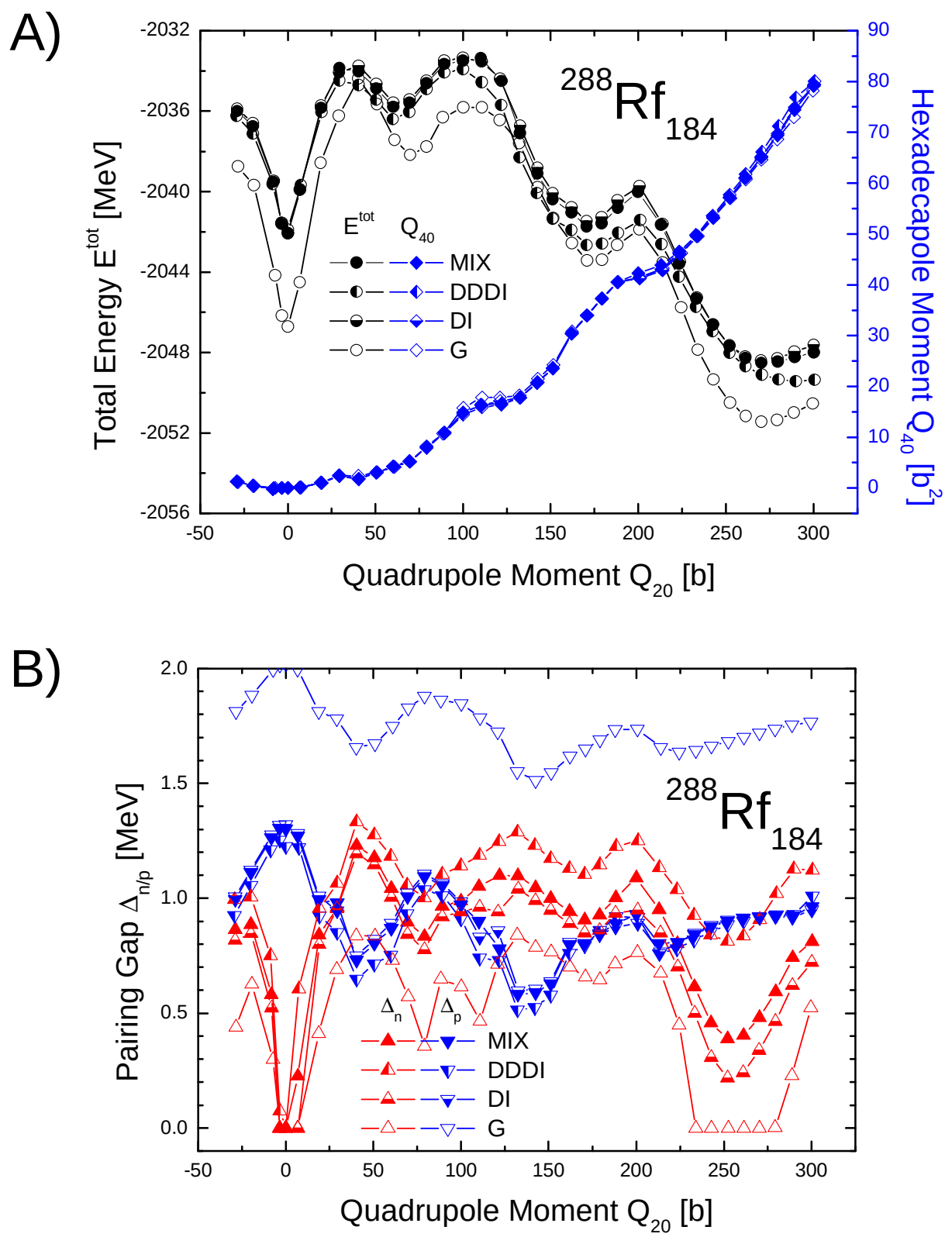

Fig. 1. (**A**) The total binding energies $E^{\rm tot}$ (left-hand side scale) and mass hexadecapole moments $Q_{40}$ (right-hand side scale) along the fission paths of $^{288}$Rf$_{184}$ calculated with the SLy4 interaction and four different pairing interactions: MIX, DDDI, and DI $\delta$-interaction, and seniority pairing (G). (**B**) The neutron $\Delta_n$ and proton $\Delta_p$ pairing gaps along the fission paths shown above.

### 2.2. *Comparison of pairing models*

Figures 1, 2, and 3 display total binding energies ($E^{\rm tot}$), mass hexadecapole moments ($Q_{40}$), and neutron/proton pairing gaps ($\Delta_{n/p}$) calculated along the static fission paths of $^{288}$Rf$_{184}$, $^{298}114_{184}$, and $^{310}126_{184}$. The fission paths were computed with a quadratic constraint[20] on the mass quadrupole moment ($Q_{20}$). Our study covers the oblate/prolate deformations of $Q_{20} = -30 \div 300$ b with a step of 10 b.

As in previous studies based on the SLy4 functional,[21,22] we found that all nuclei considered in this work have spherical shapes in their ground states. This is because $N$=184 appears to be the magic neutron number in most theoretical models based on the Skyrme approach.[23] Due to the magic character of $N$=184 isotones, ground-state neutron pairing gaps calculated in BCS vanish. The ground-state proton pairing gaps, on the other hand, show considerable variations with $Z$. They are large in the open-shell rutherfordium ($Z$=104), but in $Z$=114 the values of $\Delta_p$ are considerably reduced. As expected, proton pairing is much weakened for $Z$=126 which is predicted to be semi-magic.[21,22,23]

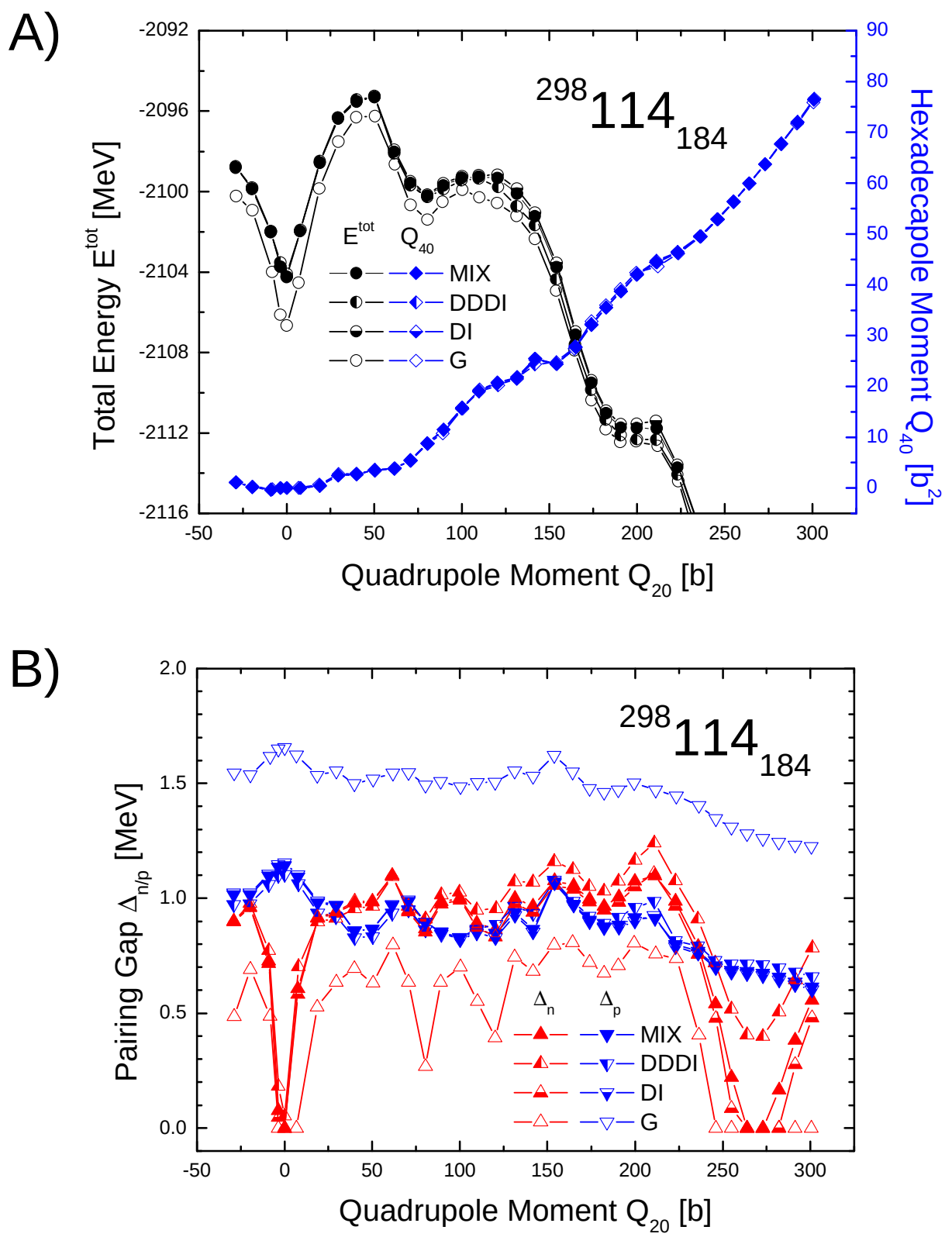

Fig. 2. Same as in Fig. 1 but now for $^{298}114_{184}$.

The calculated static fission paths have reflection-symmetric shapes, i.e., $Q_{\lambda 0} = 0$ for all odd multipolarities $\lambda$. Furthermore, one can see that irrespective of the pairing interaction used, the hexadecapole moments are almost identical, and they gradually increase (from 0 up to about 80 b$^2$) along the calculated static fission paths. In contrast to $Q_{40}$, the collective potentials (i.e., total energies $E^{\rm tot}$ as functions of $Q_{20}$) differ depending on the pairing model employed. This is particularly evident for $^{288}$Rf$_{184}$ (Fig. 1A) where the fission barrier calculated within the SHF+BCS(G) model is significantly higher as compared to those obtained in the SHF+BCS($\delta$) variants. In the case of $^{298}114_{184}$ (Fig. 2A), the fission barriers calculated with the SHF+BCS(G) and SHF+BCS($\delta$) interactions are considerably closer to one another than in the case of $^{288}$Rf$_{184}$. Furthermore, for $^{310}126_{184}$ (Fig. 3A), all fission barriers calculated with both pairing models are almost identical. This can be attributed to the behavior of proton pairing along the fission paths. Indeed, in $^{288}$Rf$_{184}$ there is a large systematic difference between $\Delta_p$ values in SHF+BCS(G) and SHF+BCS($\delta$) variants, with the seniority-pairing model producing considerably larger pairing gaps. This difference decreases when going towards $^{310}126_{184}$ in which $\Delta_{n/p}$ obtained within the SHF+BCS(G) model are much closer to those obtained within the SHF+BCS($\delta$) model. This result indicates that the isospin dependence of seniority pairing strengths given by Eq. (1) is too strong and thus unrealistic. Another interesting observation is that neutron/proton pairing gaps (hence, corresponding potential energies) calculated within the SHF+BCS($\delta$) framework are very similar, regardless of the parameterization variant used (MIX, DDDI or DI), see Eq. (4).

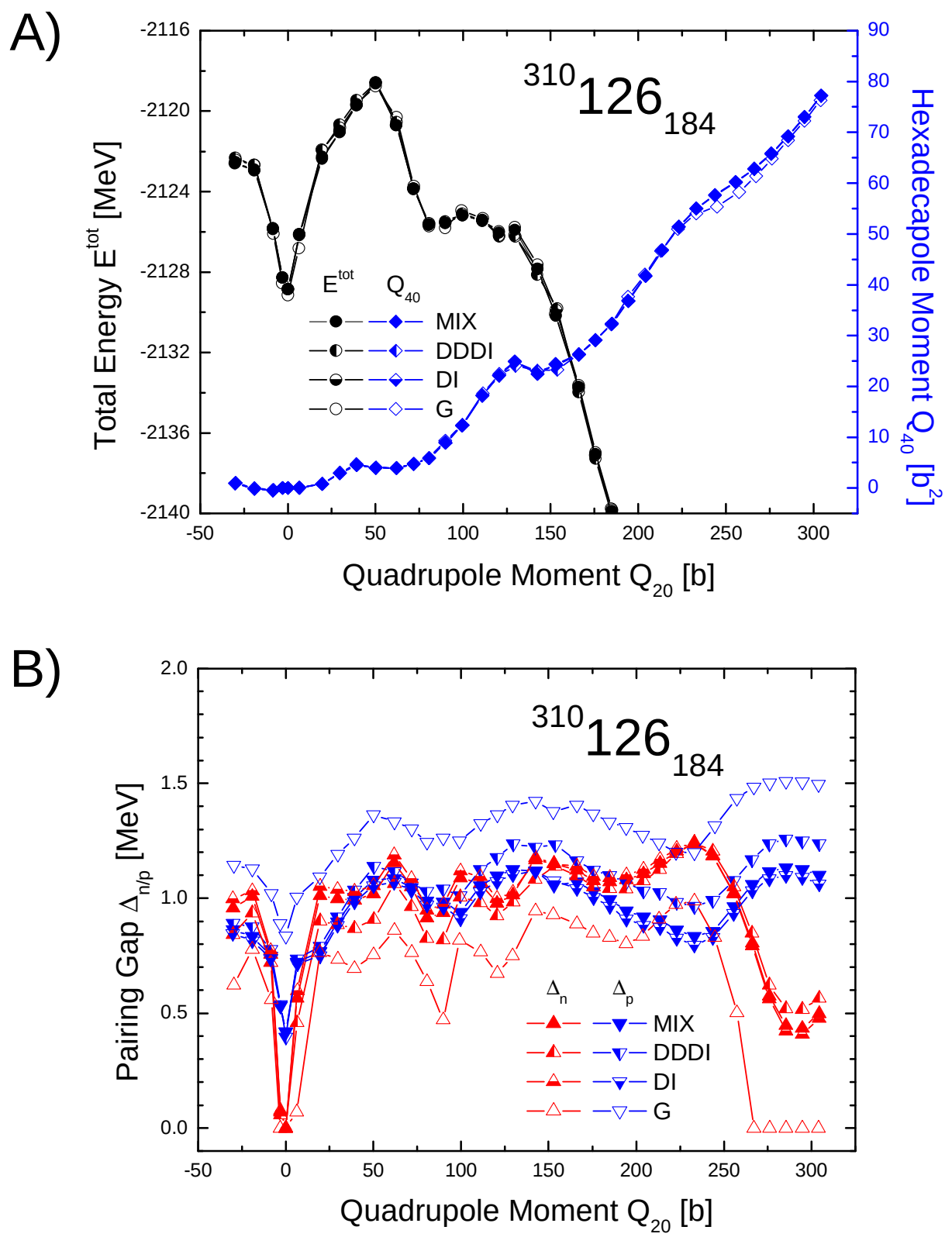

Fig. 3. Same as in Fig. 1 but now for $^{310}126_{184}$.

### 2.3. *Fission barriers for the even-even $N$=184 isotones with $\delta$-pairing interaction*

Figure 4 compares the total binding energies ($E^{\rm tot}$) and the mass hexadecapole moments ($Q_{40}$) calculated along the fission paths for twelve even-even $N$=184 isotones. Here we use the MIX parameterization of the $\delta$-interaction. We found that all of the super-heavy elements (SHE) studied here have reflection-symmetric static fis-sion paths and are spherical in the ground states, and the $Q_{40}$ moments calculated along the static fission paths follow the same pattern; i.e., their values continuously increase from 0 up to about 80 b$^2$.

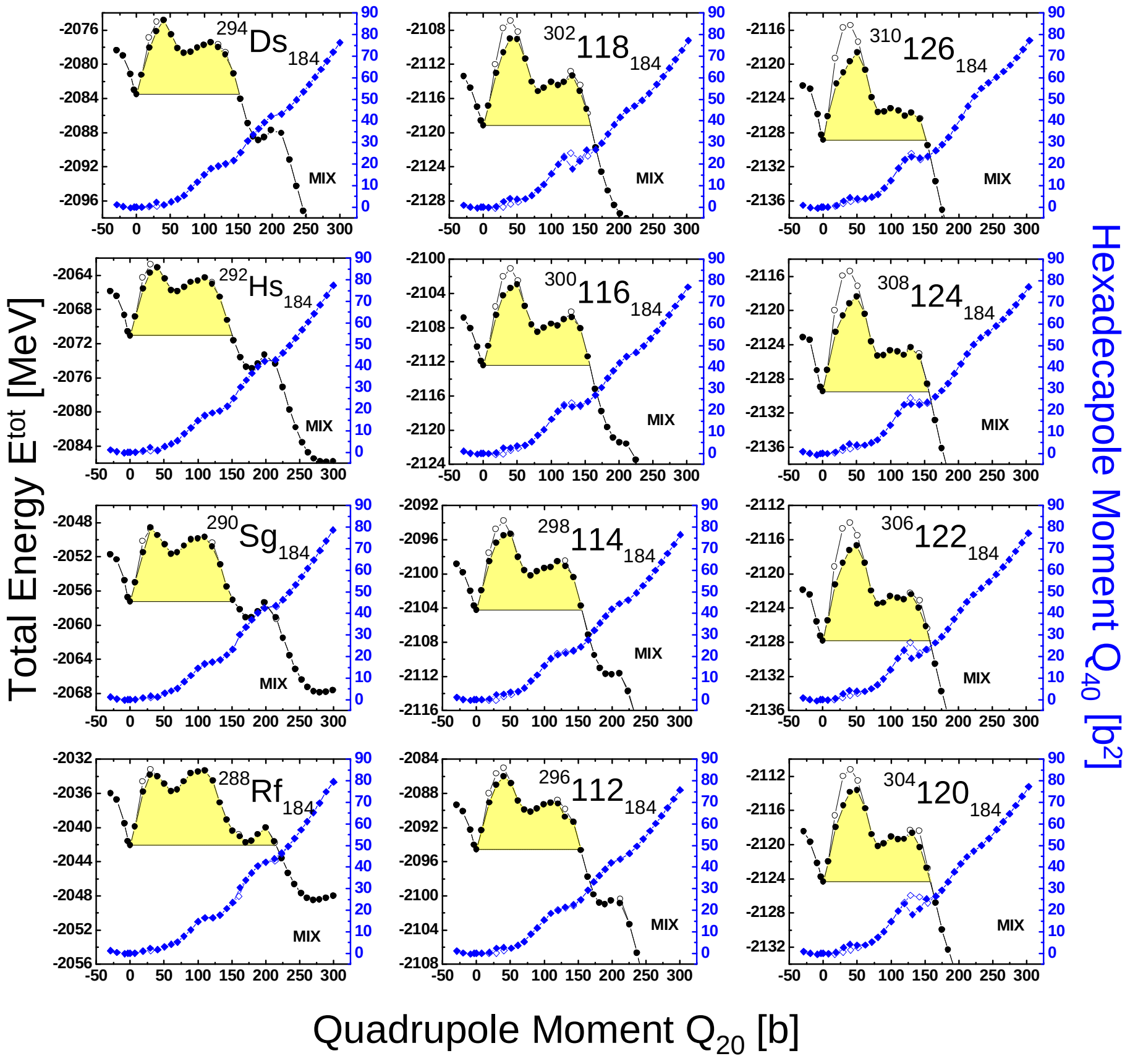

Fig. 4. The total binding energies $E^{\rm tot}$ (● - scales on the left-hand side) and mass hexadecapole moments $Q_{40}$ (♦ - scales on the right-hand side) along the fission paths, calculated with the SLy4 interaction and the MIX parameterization of the $\delta$-interaction for the even-even $N$=184 isotones of 288≤$A$≤310. The differences between open and solid symbols illustrate the effects of triaxiality on the inner barriers.

All collective potentials shown in Fig. 4 have two-humped shapes and similar widths. Only in the case of $^{288}$Rf$_{184}$, can one see an additional small third external barrier. The outer barrier is systematically reduced with $Z$. The differences between open and solid symbols illustrate the effect of triaxiality on the inner barriers. The strongest effect is predicted for the nucleus $^{310}126_{184}$, where the barrier is lowered by more than 3 MeV due to triaxiality. However, for lighter isotones with $Z \leq 114$, the role of triaxiality is modest.

For completeness, it is interesting to look at the neutron $\langle\Delta_n\rangle$ and proton $\langle\Delta_p\rangle$ spectral pairing gaps calculated along the static fission paths shown in Fig. 4. They are shown in Fig. 5. As already mentioned, the ground-state neutron pairing gaps $\langle\Delta_n\rangle$ (▲) are equal to 0 in all the $N$=184 isotones, whereas the proton gaps $\langle\Delta_p\rangle$ (▼) change from 1.3 MeV in $^{288}$Rf$_{184}$) to less than 0.5 MeV in $^{288}126_{184}$. In the sub-barrier regions of the fission paths (shadowed regions in the plot), neutron and proton spectral pairing gaps fluctuate around 1 MeV.

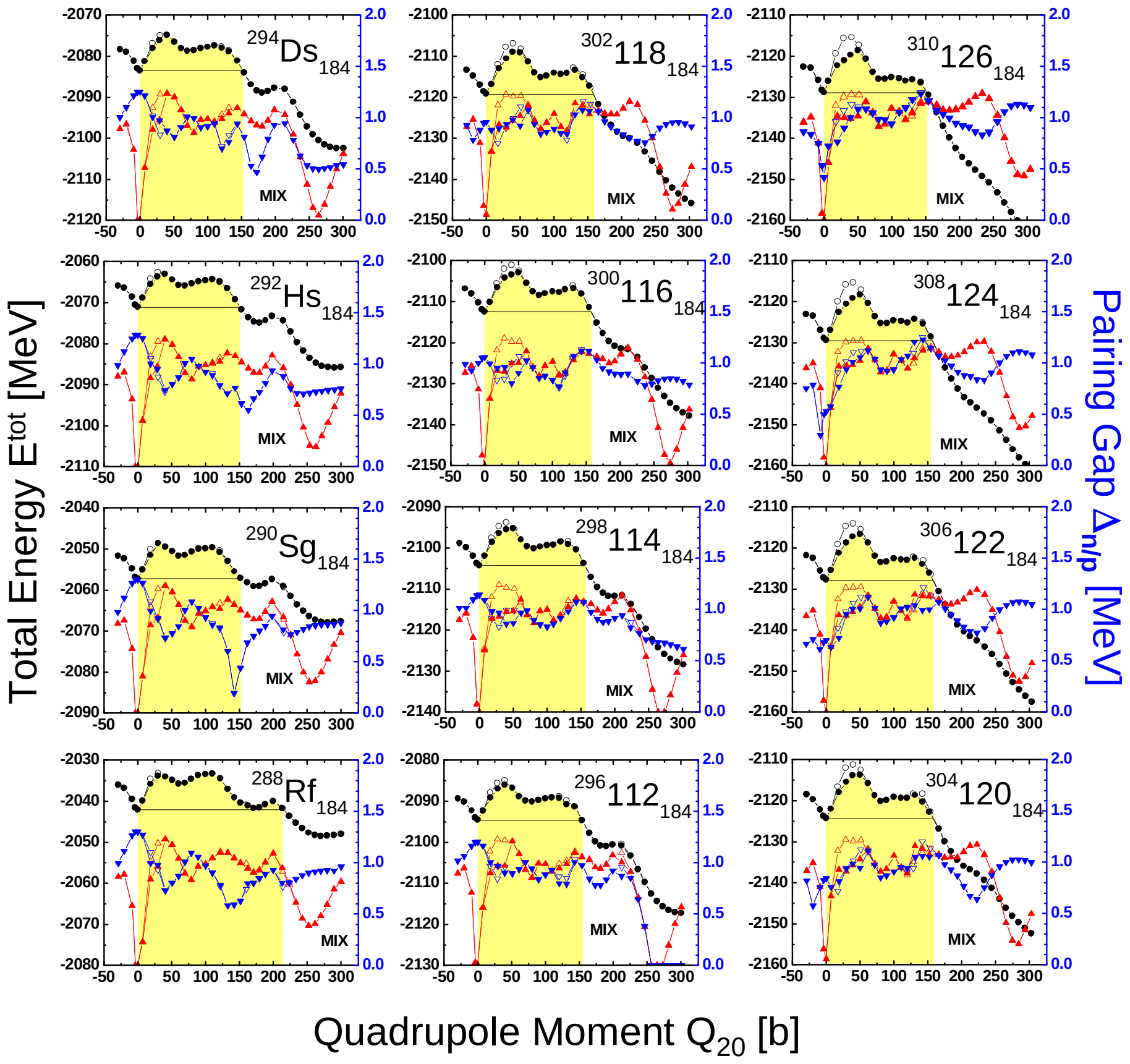

Fig. 5. The total binding energies $E^{\rm tot}$ (●, scales on the left-hand side) and the spectral neutron $\langle\Delta_n\rangle$ (▲) and proton $\langle\Delta_p\rangle$ (▼) pairing gaps (scales on the right-hand side) for the $N$=184 isotones shown in Fig. 4.

The size of a fission barrier is a measure of stability of the nucleus against spontaneous fission. The static fission barriers shown in Fig. 4 suggest a possible increase of the stability against spontaneous fission for $118 \leq Z \leq 124$. This predicted increase of the stability supports the results obtained in our previous paper,[10] where the same nuclei were considered within the SHF+BCS(G) framework, but with the pairing strengths of Eq. (1) scaled to reproduce the pairing gaps of the finite-range droplet model (FRDM).[24]

## 3. Conclusion

The purpose of this work was to study the sensitivity of static fission trajectories in $N$=184 isotones to the choice of pairing interaction in the SHF+BCS model. We found that the results (energies, hexadecapole moments, pairing gaps) are fairly insensitive to the assumed density dependence of the pairing $\delta$-interaction (DI, DDDI, or MIX). On the other hand, we found rather large differences between results obtained in $\delta$-pairing and seniority-pairing variants. This suggests that the isospin dependence of the seniority-pairing strength used is unrealistic. In general, it is always advantageous to consider a more realistic interaction in the pairing channel (such as the density-dependent $\delta$ force) which does not depend *explicitly* on mass number and neutron excess.

In the second part of the paper, we examined static fission paths calculated within the SHF+BCS($\delta$) framework and MIX pairing parameterization for twelve even-even $N$=184 isotones. We found that the reduction of fission barriers due to the appearance of triaxial deformation strongly depends on $Z$. Furthermore, the analysis of the static fission barriers suggests the increased stability against spontaneous fission for $118 \leq Z \leq 124$.

## Acknowledgements

This work was supported in part by the National Nuclear Security Administration under the Stewardship Science Academic Alliances program through the U.S. Department of Energy Research Grant DE-FG03-03NA00083; by the U.S. Department of Energy under Contract Nos. DE-FG02-96ER40963 (University of Tennessee), DE-AC05-00OR22725 with UT-Battelle, LLC (Oak Ridge National Laboratory), and DE-FG05-87ER40361 (Joint Institute for Heavy Ion Research); by the Polish Committee for Scientific Research (KBN) under Contract No. 1 P03B 059 27; by the Foundation for Polish Science (FNP); and by the Polish Ministry of Science and Higher Education under Contract No. N202 179 31/3920.